# Electron capture rates on titanium isotopes in stellar matter

**Jameel-Un Nabi[*], Muhammad Sajjad, Muneeb-Ur Rahman**

Faculty of Engineering Sciences, Ghulam Ishaq Khan Institute of Engineering Sciences and Technology, Topi 23460, Swabi, NWFP, Pakistan

Electron captures are amongst the most important weak interaction rates related to the dynamics of stellar core collapse. They play a key role in the gravitational collapse of the core of a massive star triggering the supernova explosion. Titanium isotopes are believed to have significant impact on controlling the lepton-to-baryon fraction in the late phases of evolution of core of massive stars. This work consists of the calculation of electron capture rates on titanium isotopes. The pn-QRPA theory is used to calculate electron capture rates in stellar matter. The electron capture rates are calculated over a wide range of densities ($10 \leq \rho Y_e$ (g cm$^{-3}$) $\leq 10^{11}$) and temperatures ($10^7 \leq T$ (K) $\leq 30 \times 10^9$). Here we also report the differences in electron capture rates with the earlier calculations including those using large scale shell model.

**PACS** numbers: 26.50.+x, 23.40.Bw, 23.40.-s, 21.60Jz

## 1. Introduction

The late evolution stages of massive stars are strongly influenced by weak interactions, which act to determine the core entropy and electron-to-baryon ratio ($Y_e$) of the presupernova star and hence its Chandrasekhar mass which is proportional to $Y_e^2$ [1]. Electron capture reduces the number of electrons available for pressure support, while beta decay acts in the opposite direction. Both processes generate neutrinos, which for densities less than $10^{11} g/cm^3$ escape the star, carrying away energy and entropy from the core. Electron capture and beta decay, during the final evolution of a massive star, are dominated by Fermi and Gamow-Teller (GT) transitions. While the treatment of Fermi transitions (important only in beta decays) is straightforward, a correct description of the GT transitions is a difficult problem in nuclear structure. In the astrophysical environment nuclei are fully ionized, so one has continuum electron capture from the degenerate electron plasma. The energies of the electrons are high enough to induce transitions to the GT resonance. Electron captures and beta decays occur during hydrostatic burning stages.

---

[*] Corresponding author
e-mail: jnabi00@yahoo.com
Phone: 0092-938-271858 (ext.2535),
Fax: 0092-938-271862.

They determine the physical conditions in the (quasi-) hydrostatically evolving core through their influence on the entropy per nucleon. Moreover, these processes grow in importance in later stellar evolution stages when the densities and temperatures in the core become large and the increasing Fermi energy of the electrons makes electron capture energetically favorable.

Electron captures have two important consequences in stellar evolution. First, they reduce the $Y_e$ and drive the stellar matter neutron richer. Second, the neutrinos produced leave and cool the star as the involved densities are still low enough. These very two properties make electron capture, now taking place on pf-shell nuclei in the mass range around A=60, decisive for the presupernova collapse of a massive star [2]. Stars with mass $> 8 M_\odot$ after passing through all hydrostatic burning stages develop an onion like structure; produce a collapsing core at the end of their evolution and lead to increased nuclear densities in the stellar core [3]. Electron capture on nuclei takes place in very dense environment of the stellar core where the Fermi energy (chemical potential) of the degenerate electron gas is sufficiently large to overcome the threshold energy given by negative Q values of the reactions involved in the interior of the stars. This high Fermi energy of the degenerate electron gas leads to enormous electron capture on nuclei and results in the reduction of the electron to baryon ratio.

The importance of obtaining accurate nuclear weak rates both during the hydrostatic phases of stellar evolution [4, 5] and during the collapse phase is a well known problem. Reliable electron capture rates are required to compute the dynamics of the core in fall epoch [6]. The very high density electron capture rates determine to some extent the entropy and $Y_e$ evolution in the collapsing stellar core, and hence are required for an understanding of the explosion mechanism.

The first extensive effort to tabulate the weak interaction rates at high temperatures and densities, where decays from excited states of the parent nuclei become relevant, was done by Fuller, Fowler, and Newman (FFN) [7]. FFN calculated the weak interaction rates over a wide range of temperatures and densities ($10 \leq \rho Y_e$ (g cm$^{-3}$) $\leq 10^{11}$, $10^7 \leq$ T (K) $\leq 10^{11}$). The GT strength and excitation energies were calculated using a zero-order shell model. They also incorporated the experimental data available at that time. The matrix elements of Brown et al. [8] were used for unmeasured GT strengths. When these were also not available, FFN assumed an average log *ft* value of 5.0 for their calculation of weak rates.

Later Aufderheide et al. [9] extended the FFN work for neutron rich nuclei in pf-shell. They tabulated the top 90 electron capture nuclei averaged throughout the stellar trajectory for $0.40 \leq Y_e \leq 0.5$ (see Table 25 therein). These were the nuclei which, according to the calculations of the authors, affected $\dot{Y}_e$ (rate of change of $Y_e$) the most in the presupernova evolution. Large-scale shell-model calculations have also been performed for pf-shell nuclei in the mass range A= 45-65 [10].

The list of most important electron capture nuclei (those which make the largest contribution to $\dot{Y}_e$) compiled by Aufderheide and collaborators [9] also contain the titanium isotopes of mass number 49, 51, 52, 53, and 54. Titanium isotopes have been assigned differing nucleosynthetic origins [11]. Oxygen burning for $^{46}$Ti, silicon and carbon burning for $^{47}$Ti and $^{49}$Ti, silicon burning for the abundant isotope $^{48}$Ti, and carbon burning for $^{50}$Ti [11]. This mixture of origins suggests that the titanium isotopic abundance ratios may have evolved during the life time of the Galaxy. Titanium isotopes and the commonly studied metals are both believed to be synthesized in the same nuclear-burning stages in stellar cores, the different titanium isotopes play the same interpretive role as the individual elements [11].

In type Ia SNe, which is interpreted as thermonuclear explosions of accreting white dwarf in binary stellar systems, electron capture occurs behind the burning front. Electron capture processes, dominated by $GT^+$ transitions, are not only important for the dynamics of the propagating burning front, they directly effect the composition of the ejecta from such explosions [12]. If the central density exceeds a critical value, electron capture can cause a dramatic reduction in the pressure of degenerate electrons and can therefore induce collapse (accretion-induced collapse [AIC]) of the white dwarf [13]. Thus the abundance of the Fe group, in particular of neutron rich species like $^{48}$Ca, $^{50}$Ti, $^{59}$Cr, $^{54,58}$Fe and $^{58}$Ni, is highly sensitive to the electron captures taking place in the central layers of SNe Ia explosions [14-17]. These captures drive the matter to larger neutron excesses and thus strongly influence the composition of ejected matter and the dynamics of the explosion.

In this work the electron capture rates for titanium isotopes are calculated using a microscopic theory. The proton-neutron quasi random phase approximation (pn-QRPA) theory, used in this calculation, constructs parent and daughter excited states and also calculates the Gamow–Teller strength distribution among the corresponding states. In other words the Brink's hypothesis (which states that the GT strength distribution on excited states is identical to that from the ground state, shifted only by the excitation energy of the state) is not employed in this calculation and a state-to-state calculation of the electron capture rates further increases the reliability of the calculations.

Nabi and Klapdor used the *pn*-QRPA theory to calculate weak interaction mediated rates and energy losses in stellar environment for *sd*- [18] and *fp*/*fpg*-shell nuclides [19]. Reliability of the calculated rates was also discussed in detail in [19]. There the authors compared the measured data of thousands of nuclides with the *pn*-QRPA calculations and got good comparison (see also [20]). Here we use this extended model to calculate the electron capture rates in stellar matter for titanium isotopes. Another advantage of using the *pn*-QRPA theory is that one can handle large configuration spaces. In our model, we considered a model space up to 7 major shells. We included in our calculations parent excitation energies well in excess of 10 MeV. For each parent excited state, 150 daughter excited states were considered in the vicinity of 30 MeV. The summation over all partial rates was done until satisfactory convergence was achieved.

This paper is organized as follows. In Sec. 2, we present the formalism of electron capture rate calculations. The calculated electron capture rates of titanium isotopes (A=40 to 61) and the mutual comparison with the earlier works of FFN [7] and large scale shell model calculations [10] are presented in Sec. 3. We finally conclude our reportings in Sec. 4.

## 2. Formalism

The following main assumptions are made in our calculations:

(1) Only allowed GT and super allowed Fermi transitions are calculated. It is assumed that contributions from forbidden transitions are relatively negligible.

(2) It is assumed that the temperature is high enough to ionize the atoms completely. The electrons are not bound anymore to the nucleus and obey the Fermi-Dirac distribution.

(3) The distortion of electron wave function due to the Coulomb interaction with a nucleus is represented by the Fermi function in phase space integrals.

### 2.1. Rate Formulae

The Hamiltonian of the pn-QRPA is given by

$$H^{QRPA} = H^{sp} + V^{pair} + V_{GT}^{ph} + V_{GT}^{pp}, \qquad (1)$$

where $H^{sp}$ is the single-particle Hamiltonian, $V^{pair}$ is the pairing force, $V_{GT}^{ph}$ is the particle-hole (ph) Gamow-Teller force, and $V_{GT}^{pp}$ is the particle-particle (pp) Gamow-Teller force. The Hamiltonian is diagonalized in three consecutive steps as outlined below. Single-particle energies and wave functions are calculated in the Nilsson model, which takes into account nuclear deformation [21] (more details can be found in [22]). The transformation from the spherical basis to the axial symmetric deformed basis can be written as [23]

$$d_{m\alpha}^{+} = \sum_{j} D_{j}^{m\alpha} c_{jm}^{+}, \qquad (2)$$

where $d^+$ and $c^+$ are particle creation operators in the deformed and spherical basis, respectively, the transformation matrices $D_j^{m\alpha}$ are determined by diagonalization of the Nilsson Hamiltonian, and $\alpha$ represents additional quantum numbers, except $m$, which specify the Nilsson eigenstates.

Pairing is in the *BCS* approximation, where a constant pairing force with force strength $G$ ($G_p$ and $G_n$ for protons and neutrons, respectively) is applied,

$$V^{pair} = -G \sum_{jmj'm'} (-1)^{l+j-m} c_{jm}^+ c_{j-m}^+ (-1)^{l'+j'-m'} c_{j'-m'} c_{j'm'}, \tag{3}$$

where the sum over $m$ and $m'$ is restricted to $m, m' > 0$, and $l$ donates the orbital angular momentum. The *BCS* calculation gives quasiparticle energies $\varepsilon_{m\alpha}$.

A quasiparticle basis is introduced via

$$a_{m\alpha}^+ = u_{m\alpha} d_{m\alpha}^+ - v_{m\alpha} d_{\bar{m}\alpha}, \tag{4}$$

$$a_{\bar{m}\alpha}^+ = u_{m\alpha} d_{\bar{m}\alpha}^+ + v_{m\alpha} d_{m\alpha}, \tag{5}$$

Where $\bar{m}$ is the time-reversed state of $m$, and $a^+ (a)$ are the quasiparticle creation (annihilation) operators which enter the RPA equation (the Condon-Shortley phase convention [24] is taken for the nucleon basis, and the BCS phases [**25**] for the Nilsson basis and also for quasiparticles). The occupation amplitudes u and v satisfy the condition $u^2 + v^2 = 1$ and are determined by the BCS equations (see for example [25], page 230). The ground state wave function consists mainly of the BCS ground state with no quasiparticles and the leading and the leading admixtures are four quasiparticle state.

In the pn-QRPA, charge-changing transitions are expressed in terms of phonons creation, with the QRPA phonons defined by

$$A_\omega^+(\mu) = \sum_{pn} (X_\omega^{pn}(\mu) a_p^+ a_{\bar{n}}^+ - Y_\omega^{pn}(\mu) a_n^+ a_{\bar{p}}^+). \tag{6}$$

The sum in Eq. (6) runs over all proton-neutron pairs with $\mu = m_p - m_n = -1, 0, 1$ where $m_{p(n)}$ denotes the third component of the angular momentum. The ground state of the theory is defined as the vacuum with respect to the QRPA phonons, $A_\omega(\mu)|QRPA\rangle = 0$. The forward- and backward-going amplitudes X and Y are eigenfunctions of the RPA matrix equation

$$\begin{bmatrix} A & B \\ -B & -A \end{bmatrix} \begin{bmatrix} X \\ Y \end{bmatrix} = \omega \begin{bmatrix} X \\ Y \end{bmatrix}, \tag{7}$$

where $\omega$ are energy eigenvalues of the eigenstates and elements of the two sub-matrices are are given by

$$A_{pn,p'n'} = \delta(pn, p'n')(\varepsilon_p + \varepsilon_n) + V^{pp}_{pn,p'n'}(u_p u_n u_{p'} u_{n'} + v_p v_n v_{p'} v_{n'}) +$$
$$V^{ph}_{pn',p'n'}(u_p v_n u_{p'} v_{n'} + v_p u_n v_{p'} u_{n'}), \tag{8}$$

$$B_{pn,p'n'} = V^{pp}_{pn,p'n'}(u_p u_n v_{p'} v_{n'} + v_p v_n u_{p'} u_{n'}) - V^{ph}_{pn,p'n'}(u_p v_n v_{p'} u_{n'} + v_p u_n u_{p'} v_{n'}). \tag{9}$$

The backward–going amplitude Y accounts for the ground-state correlations.

In the present work, in addition to the well known particle-hole force [26, 27]

$$V^{ph}_{GT} = 2\chi \sum_\mu (-1)^\mu Y_\mu Y^+_{-\mu}, \tag{10}$$

with

$$Y_\mu = \sum_{j_p m_p j_n m_n} <j_p m_p | t_-\sigma_\mu | j_n m_n> c^+_{j_p m_p} c_{j_n m_n}, \tag{11}$$

The particle-particle interaction, approximated by the separable force [28, 29]

$$V^{pp}_{GT} = -2\kappa \sum_\mu (-1)^\mu P^+_\mu P_{-\mu}, \tag{12}$$

with

$$P^+_\mu = \sum_{j_p m_p j_n m_n} <j_n m_n | (t_-\sigma_\mu)^+ | j_p m_p> (-1)^{l_n+j_n-m_n} c^+_{j_p m_p} c^+_{j_n-m_n}, \tag{13}$$

is taken into account. The interaction constants $\chi$ and $\kappa$ in units of MeV are both taken to be positive.

Matrix elements of the forces which appear in RPA equation (7) are separable,

$$V^{ph}_{pn,p'n'} = +2\chi f_{pn}(\mu) f_{p'n'}(\mu), \tag{14}$$

$$V^{pp}_{pn,p'n'} = -2\kappa f_{pn}(\mu) f_{p'n'}(\mu), \tag{15}$$

with
$$f_{pn}(\mu) = \sum_{j_p j_n} D^{m_p \alpha_p}_{j_p} D^{m_n \alpha_n}_{j_n} < j_p m_p | t_- \sigma_\mu | j_n m_n >, \tag{16}$$

which are single-particle Gamow-teller transition amplitudes defined in the Nilsson basis. For the separable forces, the matrix equation (7) can be expressed more explicitly,

$$X^{pn}_\omega = \frac{1}{\omega - \varepsilon_{pn}} [2\chi(q_{pn} Z^-_\omega + q_{pn} Z^+_\omega) - 2\kappa(q^U_{pn} Z^{--}_\omega + q^V_{pn} Z^{++}_\omega)], \tag{17}$$

$$Y^{pn}_\omega = \frac{1}{\omega + \varepsilon_{pn}} [2\chi(q_{pn} Z^+_\omega + q_{pn} Z^-_\omega) + 2\kappa(q^U_{pn} Z^{++}_\omega + q^V_{pn} Z^{--}_\omega)], \tag{18}$$

where $\varepsilon_{pn} = \varepsilon_p + \varepsilon_n$, $\quad q_{pn} = f_{pn} u_p v_n$, $\quad q^U_{pn} = f_{pn} u_p u_n$, $\quad q_{pn} = f_{pn} v_p u_n$, $\quad q^V_{pn} = f_{pn} v_p v_n$,

$$Z^-_\omega = \sum_{pn}(X^{pn}_\omega q_{pn} - Y^{pn}_\omega q_{pn}), \tag{19}$$

$$Z^+_\omega = \sum_{pn}(X^{pn}_\omega q_{pn} - Y^{pn}_\omega q_{pn}), \tag{20}$$

$$Z^{--}_\omega = \sum_{pn}(X^{pn}_\omega q^U_{pn} + Y^{pn}_\omega q^V_{pn}), \tag{21}$$

$$Z^{++}_\omega = \sum_{pn}(X^{pn}_\omega q^V_{pn} + Y^{pn}_\omega q^U_{pn}). \tag{22}$$

It is noted that the right –hand-side of (17) and (18) involves $X^{pn}_\omega$ and $Y^{pn}_\omega$ through $Z$ of (22). Inserting (17) and (18) into (22), $X^{pn}_\omega$ and $Y^{pn}_\omega$ can be eliminated which explicitly depend on individual proton-neutron quasiparticle pairs. One then obtains a set of equations for $Z^-$, $Z^+$, $Z^{--}$ and $Z^{++}$, which are equivalent to the matrix equation (7),

$$Mz = 0, \tag{23}$$

where
$$M = \begin{bmatrix} \chi M_1 - 1 & \chi M_0 & -\kappa M_5 & -\kappa M_7 \\ \chi M_0 & \chi M_2 - 1 & -\kappa M_8 & -\kappa M_6 \\ \chi M_5 & \chi M_8 & -\kappa M_3 - 1 & -\kappa M_0 \\ \chi M_7 & \chi M_6 & -\kappa M_0 & -\kappa M_4 - 1 \end{bmatrix}, \quad (24)$$

$$z = \begin{bmatrix} Z_\omega^- \\ Z_\omega^+ \\ Z_\omega^{--} \\ Z_\omega^{++} \end{bmatrix},$$

(25)

and

$$M_0 = 2\sum_{pn}\left(\frac{q_{pn}q_{pn}}{\omega - \varepsilon_{pn}} - \frac{q_{pn}q_{pn}}{\omega + \varepsilon_{pn}}\right) = 2\sum_{pn}\left(\frac{q_{pn}^U q_{pn}^V}{\omega - \varepsilon_{pn}} - \frac{q_{pn}^U q_{pn}^V}{\omega + \varepsilon_{pn}}\right),$$

$$M_1 = 2\sum_{pn}\left(\frac{q_{pn}^2}{\omega - \varepsilon_{pn}} - \frac{q_{pn}^2}{\omega + \varepsilon_{pn}}\right),$$

$$M_2 = 2\sum_{pn}\left(\frac{q_{pn}^2}{\omega - \varepsilon_{pn}} - \frac{q_{pn}^2}{\omega + \varepsilon_{pn}}\right),$$

$$M_3 = 2\sum_{pn}\left(\frac{q_{pn}^{U^2}}{\omega - \varepsilon_{pn}} - \frac{q_{pn}^{V^2}}{\omega + \varepsilon_{pn}}\right),$$

$$M_4 = 2\sum_{pn}\left(\frac{q_{pn}^{V^2}}{\omega - \varepsilon_{pn}} - \frac{q_{pn}^{U^2}}{\omega + \varepsilon_{pn}}\right),$$

$$M_5 = 2\sum_{pn}\left(\frac{q_{pn}q_{pn}^U}{\omega - \varepsilon_{pn}} - \frac{q_{pn}q_{pn}^V}{\omega + \varepsilon_{pn}}\right),$$

$$M_6 = 2\sum_{pn}\left(\frac{q_{pn}q_{pn}^V}{\omega - \varepsilon_{pn}} - \frac{q_{pn}q_{pn}^U}{\omega + \varepsilon_{pn}}\right),$$

$$M_7 = 2\sum_{pn}\left(\frac{q_{pn}q_{pn}^V}{\omega-\varepsilon_{pn}} - \frac{q_{pn}q_{pn}^U}{\omega+\varepsilon_{pn}}\right),$$

$$M_8 = 2\sum_{pn}\left(\frac{q_{pn}q_{pn}^U}{\omega-\varepsilon_{pn}} - \frac{q_{pn}q_{pn}^V}{\omega+\varepsilon_{pn}}\right), \quad (26)$$

Equation (23) has a solution, when the determinant of matrix M vanishes,

$$\det M = 0, \quad (27)$$

by regarding $M_k(k=0-8)$ as functions of the energy variable $\omega$. Thus, the eigenvalue problem of (7) is reduced to finding roots of the algebraic equation (27). Equation (27) is simplified to an equation of second order in $M_k$'s for no pp force $(\kappa=0)$ and with the inclusion of pp force it is an equation of fourth order. The solution of this equation is not difficult and can be found in [30].

For each eigenvalue $\omega$, Gamow-Teller transition amplitudes to the RPA eigenstate are calculated as follows. First, four co determinants $N_\omega^-, N_\omega^+, N_\omega^{--}$ and $N_\omega^{++}$ of $M$ are evaluated at the eigenvalue $\omega$. The co determinants are obtained by expanding $\det M$ with respect to the first row,

$$\det M = (\chi M_1 - 1)N^- + \chi M_0 N^+ - \kappa M_5 N^{--} - \kappa M_7 N^{++}. \quad (28)$$

Then, ratios of $Z_\omega$'s are calculated by

$$\frac{Z_\omega^-}{N_\omega^-} = \frac{Z_\omega^+}{N_\omega^+} = \frac{Z_\omega^{--}}{N_\omega^{--}} = \frac{Z_\omega^{++}}{N_\omega^{++}}, \quad (29)$$

and the absolute values are determined by the normalization condition of the phonon amplitudes

$$\sum_{pn}\left[\left(X_\omega^{pn}\right)^2 - \left(Y_\omega^{pn}\right)^2\right] = 1, \quad (30)$$

by inserting $Z_\omega$'s into (17) and (18). Gamow-Teller transition amplitudes from the QRPA ground state $|-\rangle$ (QRPA vacuum; $A_\omega(\mu)|-\rangle = 0$) to one-phonon states $|\omega, \mu\rangle = A_\omega^+(\mu)|-\rangle$ are readily calculated,

$$\langle \omega, \mu | t_\pm \sigma_\mu | - \rangle = \mp Z_\omega^\pm. \tag{31}$$

Excitation energies of the one-phonon states are given by $\omega - (\varepsilon_p + \varepsilon_n)$, where $\varepsilon_p$ and $\varepsilon_n$ are energies of the single quasiparticle states of the smallest quasiparticle energy in the proton and neutron systems, respectively.

The weak decay rate from the *ith* state of the parent to the *jth* state of the daughter nucleus is given by [1]

$$\lambda_{ij} = \ln 2 \frac{f_{ij}(T, \rho, E_f)}{(ft)_{ij}}, \tag{32}$$

where $(ft)_{ij}$ is related to the reduced transition probability $B_{ij}$ of the nuclear transition by

$$(ft)_{ij} = D/B_{ij}. \tag{33}$$

The D appearing in Eq. (33) is compound expression of physical constants,

$$D = \frac{2\ln 2 \hbar^7 \pi^3}{g_V^2 m_e^5 c^4}, \tag{34}$$

and,

$$B_{ij} = B(F)_{ij} + \left(g_A/g_V\right)^2 B(GT)_{ij}, \tag{35}$$

where B (F) and B (GT) are reduced transition probabilities of the Fermi and Gamow-Teller (GT) transitions, respectively,

$$B(F)_{ij} = \frac{1}{2J_i + 1} |< j \| \sum_k t_\pm^k \| i >|^2, \tag{36}$$

$$B(GT)_{ij} = \frac{1}{2J_i + 1} |< j \| \sum_k t_\pm^k \vec{\sigma}^k \| i >|^2. \tag{37}$$

In Eq. (37), $\vec{\sigma}^k$ is the spin operator and $t_\pm^k$ stands for the isospin raising and lowering operator. The value of D=6295 s is adopted and the ratio of the axial-vector ($g_A$) to the vector $(g_V)$ coupling constant is taken as 1.254. For the calculation of nuclear matrix elements we refer to [18].

---
[1] Throughout natural units ($\hbar = c = m_e = 1$) are adopted, unless otherwise stated, where $m_e$ is the electron mass.

The phase space integral $(f_{ij})$ is an integral over total energy,

$$f_{ij} = \int_{w_l}^{\infty} w\sqrt{w^2-1}(w_m+w)^2 F(+Z,w) G_- dw. \qquad (38)$$

In Eq. (38), $w$ is the total kinetic energy of the electron including its rest mass, $w_l$ is the total capture threshold energy (rest + kinetic) for electron capture. One should note that if the corresponding electron emission total energy, $w_m$ is greater than -1 then $w_l = 1$, and if it is less than or equal to 1, then $w_l = |w_m|$. $w_m$ is the total $\beta$-decay energy,

$$w_m = m_p - m_d + E_i - E_j, \qquad (39)$$

where $m_p$ and $E_i$ are mass and excitation energies of the parent nucleus, and $m_d$ and $E_i$ of the daughter nucleus, respectively.

$G_-$ is the electron distribution function. Assuming that the electrons are not in a bound state, this is the Fermi-Dirac distribution function,

$$G_- = [\exp(\frac{E - E_f}{kT}) + 1]^{-1}, \qquad (40)$$

here $E = (w-1)$ is the kinetic energy of the electrons, $E_f$ is the Fermi energy of the electrons, T is the temperature, and $k$ is the Boltzmann constant.

In the calculations, the inhibition of the final neutrino phase space is never appreciable enough that neutrino (or anti-neutrino) distribution functions had to be taken into consideration. $F(+Z,w)$ is the Fermi function and are calculated according to the procedure adopted by Gove and Martin [31].

The number density of electrons associated with protons and nuclei are $\rho Y_e N_A$, where $\rho$ is the baryon density, $Y_e$ is the ratio of electron number to the baryon number, and $N_A$ is the Avogadro's number.

$$\rho Y_e = \frac{1}{\pi^2 N_A} (\frac{m_e c}{\hbar})^3 \int_0^{\infty} (G_- - G_+) p^2 dp, \qquad (41)$$

where $p = (w^2-1)^{1/2}$ the electron momentum, and Eq. (41) has the units of *moles cm*$^{-3}$. This equation is used for an iterative calculation of Fermi energies for selected values of $\rho Y_e$ and $T$. $G_+$ is the positron distribution function, the Fermi-Dirac distribution function,

$$G_+ = [\exp(\frac{E+2+E_f}{kT})+1]^{-1}. \qquad (42)$$

There is a finite probability of occupation of parent excited states in the stellar environment as a result of the high temperature in the interior of massive stars. Weak interaction rates then also have a finite contribution from these excited states. The occupation probability of a state $i$ is calculated on the assumption of thermal equilibrium,

$$P_i = \frac{(2J_i+1)\exp(-E_i/kT)}{\sum_{i=1}(2J_i+1)\exp(-E_i/kT)}, \qquad (43)$$

where $J_i$ and $E_i$ are the angular momentum and excitation energy of the state $i$, respectively.

Unfortunately one cannot calculate the $J_i's$ in QRPA theory. Eq. (43) is modified as

$$P_i = \frac{\exp(-E_i/kT)}{\sum_{i=1}\exp(-E_i/kT)}. \qquad (44)$$

This approximation is a compromise and can be justified when one takes into consideration the uncertainty in the calculation of $E_i$ which over-sheds the uncertainty in calculating the values of $J_i$ in the above Eq. (43).

The rate per unit time per nucleus for electron capture, $\lambda_{ec}$ is finally given by

$$\lambda_{ec} = \sum_{ij} P_i \lambda_{ij}. \qquad (45)$$

The summation over all initial and final states is carried out until satisfactory convergence in the rate calculations is achieved.

## 3. Results and discussion

In this work we have calculated the electron capture rates for 22 isotopes of titanium. The mass range covered is from A=40 to A=61. These include the stable isotopes of titanium ($^{46}$Ti - $^{50}$Ti), as well as unstable isotopes (including neutron rich isotopes). The calculated electron capture rates are shown on an abbreviated temperature –density scale, in Table I. We present these electron capture rates for densities in the range ($10^3$-$10^{11}$gcm$^{-3}$) and at selected temperatures (1.0, 3.0, and 10.0) $\times 10^9$K. It is to be noted that the calculated electron capture rates [Eq. (45)] are tabulated in $\log_{10}\lambda_{ec}$ (in units of sec$^{-1}$). As expected, the electron capture rates increase with increasing density and temperature. The rate of change of electron capture with temperature decreases at high densities. The complete set of electron capture rates for the 22 nuclides of titanium can be obtained as files from the authors on request.

We present the comparison of our calculated Gamow-Teller strength function for $^{48}$Ti with those of measured [32] and shell model calculations [33] in Fig.1. GT transition is important for the calculations of capture rates. As can be seen in Fig.1, our GT strength of 0.91 units distributed between 2-5 MeV is comparable with experiment, which is 0.54 units in the same energy region. Authors in [33] show little strength in this region and predict a concentration of strength at higher excitation energies (4-7 MeV). Our total GT strength of 1.12 is in reasonable agreement with experimental value of 1.43 [32]. Our calculated GT strength is some what higher than the corresponding shell model result of 1.04 [33]. Concentration of the GT strength of our calculation lies in the low energy region (in the vicinity of 2.5 MeV). This is to be contrasted with shell model results where the bulk of the strength lies around 6 MeV. The shell model result [33] has been renormalized by 0.6.

In our calculation of the Gamow –Teller transitions, a quenching of the transitions was not explicitly taken into account. The quenching of the Gamow–Teller strength cannot be a constant renormalization of the axial vector current [34, 35]. We did not choose a global quenching factor (arising from higher-order configuration mixing and adopted in many shell-model calculations) for the following reasons. In order to reproduce the Gamow–Teller strength, the Gamow–Teller interaction strength parameters $\chi$ (for particle–hole interactions) and $\kappa$ (for particle–particle interactions) were adjusted in our calculations. These values were deduced from a fit to experimental half-lives. With the large model space (seven major shells) considered in this calculation and the fine tuning of GT strength parameter, we found no appreciable improvement in comparison of our calculations with the experiments by incorporating an extra quenching factor. (For a detailed discussion see also [36]).

Panels (a), (b), (c), (d) and (e) of Fig. 2 show the calculated electron capture rates of some important titanium isotopes ($^{49}$Ti, $^{51}$Ti , $^{52}$Ti, $^{53}$Ti, and $^{54}$Ti) also included in the Aufderheide's list of key nuclides [9]. These graphs depict that in a low density region, the electron capture rates remain, more or less, constant for temperatures (1.0, 3.0, and 10.0) ×10$^9$K. However as the density increases, the Fermi energy of the electron increases, and result in the enhancement of the electron capture rates in high density region (10$^8$ g-cm$^{-3}$-10$^{11}$ g-cm$^{-3}$). The region of constant electron capture rates, in these figures, with increasing temperature, shows that before core collapse the beta-decay competes with electron capture rate. At the later stages of the collapse, beta-decay becomes unimportant as an increased electron chemical potential, which grows like $\rho^{1/3}$ during in fall, drastically reduces the phase space. These results in increased electron capture rates during the collapse making the matter composition more neutron-rich. Beta-decay is thus rather unimportant during the collapse phase due to the Pauli-blocking of the electron phase space in the final state [37].

Figures (3-5) show the comparison of our work with the earlier works of FFN and shell-model rates [10]. The comparison is done wherever data for comparison was available. In comparing the electron capture rates, we depict three graphs for each isotope. The upper graph is at $1.0\times10^9$ K, the middle and lower is at $3.0\times10^9$ K and $10.0\times10^9$ K, respectively. We did this comparison for three selected values of densities i.e. $10^3$ g-cm$^{-3}$, $10^7$ g-cm$^{-3}$, and $10^{11}$ g-cm$^{-3}$. Figures 3, 4 and 5 show the comparison for $^{46,47}$Ti, $^{48,49}$Ti and $^{50,51}$Ti, respectively. Fig. 3 (a,b) and Fig. 4 (b) show the comparison of electron capture rates for nuclei $^{46}$Ti, $^{47}$Ti and $^{49}$Ti, respectively. These figures follow similar trends and one sees that pn-QRPA rates (our rates) are enhanced than FFN [7] and shell model rates [10] for all values of temperature and densities. Our rates are stronger and enhanced by as much as one to two orders of magnitude for the isotopic chain of titanium at these different selected temperatures and densities.

Fig. 5 (a,b) shows the comparison of electron capture rates for nuclei $^{50}$Ti and $^{51}$Ti, respectively. In these figures we see that the agreement is good at lower temperature ($1.00\times10^9$ K) and then at higher temperatures ($3.00\times10^9$ K and $10\times10^9$ K), where the occupation probability of excited states is greater, our rates are again one to two orders of magnitude more than those of FFN [7] and shell model rates [10].

We note that for the case of $^{48}$Ti (see Fig. 4(a)) the FFN rates are enhanced as compared to our rates in the low temperature and density regions. However at higher temperatures our rates are again enhanced (by as much as a factor of two). Our comparison with the shell model rates shows that we are in reasonable agreement at lower temperatures and enhanced at higher temperatures (again by a factor of two).

There are several reasons for the enhancement of our rates. The calculation for titanium isotopes is done in a large model space of upto seven major shells. The inclusion of a model space of $7\hbar\omega$ provided enough space to adequately handle excited states in parent and daughter nuclei. Since there is finite probability of occupation of parent excited states in the stellar environment and the transitions from these states have important contributions to the weak rates. We also do not assume the Brink's hypothesis in our rate calculations to estimate the contribution from parent excited states. In previous compilations, the transitions from excited states of parent are either ignored due to complexity of the problem or the Brink's hypothesis is applied while taking these transitions into considerations. Instead, we performed a state-by-state evaluation of electron capture rates from parent to daughter in a microscopic fashion and summed them at the end to get the total electron capture rates.

We did compare our B(GT) strength function for $^{48}$Ti with experiment [32] and shell model results [33]. We found satisfactory agreement with [32], but we extracted more strength as compared to shell model results [33] in low energy. This low lying strength contributes effectively in low density region of the stellar core. The electron capture rates are sensitive to the details of GT strength in low density region of the stellar core, which also results in the enhancement of our rates at higher temperature. In the domain of high density region the Fermi energies of electrons are high enough and the electron capture rates are sensitive to the total GT strength rather than its distribution details [28,38]. Our total GT strength is little more in comparison with the shell model and contributes in enhancement of our rates in high density region (see Fig. 4(a)). To the best of our knowledge we were not able to find similar data for comparison of GT strength distribution for other isotopes of titanium.

# 4. Conclusions

Here in this paper, we present an alternative approach to large scale shell model, i.e. pn-QRPA for the calculation of electron capture rates. Both microscopic theories use their own parameters. Both the microscopic approaches have their own associated pros and cons. However the bottom line of the comparison is that QRPA rates are enhanced in the presupernova epoch and this is a curious finding. Collapse simulators may like to find the effect of our enhanced capture rates in their stellar codes. Credibility of the weak rates is a key issue and of decisive importance for simulation codes. The credibility of the pn-QRPA has already been established [19, 20, 39]. There the authors compared the measured data (half lives and B(GT) strength) of thousands of nuclides with the pn-QRPA calculations and got fairly good comparison. Only then the theory was employed for the calculation of electron capture rates for titanium isotopes.

From astrophysical point of view these enhanced rates might have consequential effect on the late stage evolution of massive stars and the energetics of the shock waves. Results of simulations show that electron capture rates have a strong impact on the trajectory of core collapse and the properties of the core at bounce. One can not conclude just on the basis of one kind of nucleus about the dynamics of explosions (prompt or delayed). We recall that it is the rate *and* abundance of particular specie of nucleus that prioritizes the importance of that particular nucleus in controlling the dynamics of late stages of stellar evolution. The spherically symmetric core collapse simulations, taking into consideration electron capture rates on heavy nuclides, still do not explode because of the reduced electron capture in the outer layers slowing the collapse and resulting in a shock radius of slightly larger magnitude [40]. We are working on other interesting isotopes of Fe-group nuclei which play a vital role in presupernova and supernova conditions. We hope to report soon on other important electron capture/beta decay rates and associated GT strength distributions of Fe-group nuclei. It will then be interesting to find the effects of incorporating our electron capture rates on heavy nuclides in the spherically symmetric core collapse simulations.

## TABLE I

Calculated electron capture rates on titanium isotopes ($^{40}$Ti - $^{61}$Ti) for different selected densities and temperatures in stellar matter. log $\rho Y_e$ has the units of g-cm$^{-3}$, where $\rho$ is the baryon density and $Y_e$ is the ratio of the electron number to the baryon number. Temperatures ($T_9$) are measured in $10^9$ K. The calculated electron capture rates are tabulated in $\log_{10}\lambda_{ec}$ in units of sec$^{-1}$. In the table, -100 means that the rate (or the probability) is smaller than $10^{-100}$.

| log $\rho Y_e$ | $T_9$ | $^{40}$Ti | $^{41}$Ti | $^{42}$Ti | $^{43}$Ti | $^{44}$Ti | $^{45}$Ti | $^{46}$Ti | $^{47}$Ti | $^{48}$Ti | $^{49}$Ti | $^{50}$Ti |
|---|---|---|---|---|---|---|---|---|---|---|---|---|
| 3  | 1  | -3.272 | -3.569 | -3.720 | -4.213 | -10.093 | -6.318  | -18.691 | -11.050 | -27.070 | -16.927 | -40.870 |
| 3  | 3  | -0.960 | -1.258 | -1.400 | -1.944 | -6.882  | -3.922  | -7.597  | -5.642  | -9.878  | -7.220  | -14.263 |
| 3  | 10 | 0.982  | 0.771  | 0.610  | 0.084  | -1.229  | -1.126  | -1.799  | -1.734  | -2.177  | -2.160  | -3.230  |
| 7  | 1  | 0.444  | 0.153  | 0.002  | -0.472 | -5.652  | -2.509  | -12.871 | -5.305  | -21.389 | -11.107 | -35.050 |
| 7  | 3  | 0.461  | 0.166  | 0.024  | -0.507 | -5.271  | -2.457  | -5.893  | -3.944  | -8.185  | -5.506  | -12.555 |
| 7  | 10 | 1.074  | 0.864  | 0.703  | 0.178  | -1.135  | -1.030  | -1.703  | -1.637  | -2.079  | -2.061  | -3.132  |
| 11 | 1  | 5.431  | 5.293  | 5.201  | 5.046  | 4.419   | 4.528   | 4.392   | 4.564   | 4.455   | 4.430   | 4.031   |
| 11 | 3  | 5.432  | 5.289  | 5.202  | 5.039  | 4.420   | 4.531   | 4.374   | 4.571   | 4.456   | 4.443   | 4.012   |
| 11 | 10 | 5.443  | 5.345  | 5.234  | 5.068  | 4.518   | 4.616   | 4.321   | 4.602   | 4.474   | 4.504   | 4.007   |

## TABLE I (continued)

| log $\rho Y_e$ | $T_9$ | $^{51}$Ti | $^{52}$Ti | $^{53}$Ti | $^{54}$Ti | $^{55}$Ti | $^{56}$Ti | $^{57}$Ti | $^{58}$Ti | $^{59}$Ti | $^{60}$Ti | $^{61}$Ti |
|---|---|---|---|---|---|---|---|---|---|---|---|---|
| 3  | 1  | -38.852 | -51.657 | -49.359 | -62.079 | -61.701 | -73.198 | -70.022 | -83.227 | -81.336 | -100    | -89.666 |
| 3  | 3  | -13.477 | -17.708 | -16.556 | -20.335 | -20.025 | -23.820 | -22.449 | -26.988 | -26.150 | -33.046 | -30.247 |
| 3  | 10 | -2.968  | -3.741  | -3.423  | -4.068  | -3.955  | -4.901  | -4.379  | -5.915  | -5.287  | -8.137  | -7.995  |
| 7  | 1  | -33.032 | -45.837 | -43.539 | -56.259 | -55.881 | -67.378 | -64.202 | -77.407 | -75.516 | -94.372 | -83.846 |
| 7  | 3  | -11.762 | -15.993 | -14.840 | -18.619 | -18.310 | -22.105 | -20.734 | -25.273 | -24.435 | -31.331 | -28.532 |
| 7  | 10 | -2.870  | -3.642  | -3.324  | -3.969  | -3.856  | -4.802  | -4.280  | -5.816  | -5.188  | -8.038  | -7.896  |
| 11 | 1  | 3.630   | 3.817   | 3.471   | 3.639   | 3.172   | 3.113   | 2.947   | 2.807   | 2.517   | 1.979   | 2.283   |
| 11 | 3  | 3.899   | 3.812   | 3.474   | 3.641   | 3.153   | 3.117   | 2.974   | 2.812   | 2.535   | 1.993   | 2.396   |
| 11 | 10 | 4.356   | 3.871   | 3.997   | 3.804   | 3.543   | 3.324   | 3.333   | 3.047   | 2.904   | 2.425   | 2.688   |

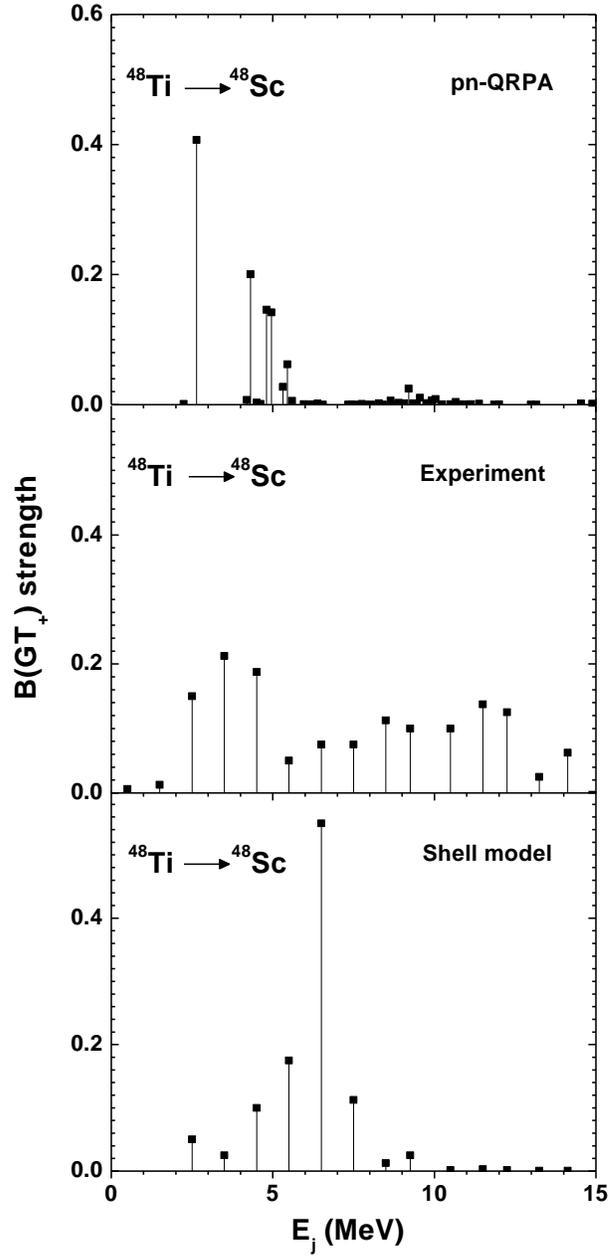

**Fig. 1.** Gamow-Teller (GT) strength distributions for $^{48}$Ti. The upper panel shows our results of GT strength for the ground state. The middle and lower panels show the results for the corresponding measured values [22] and shell model [23], respectively. $E_j$ represents daughter states in $^{48}$Sc.

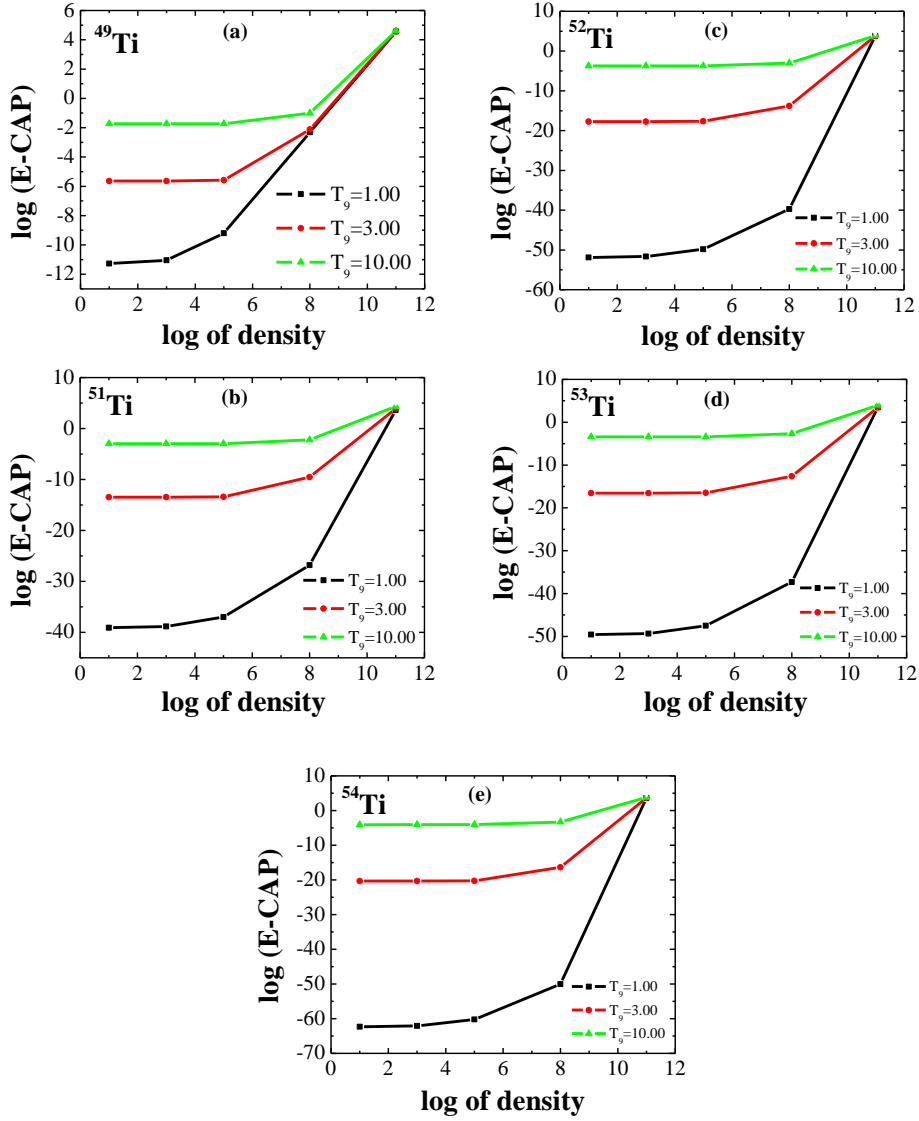

**Fig. 2.** Electron capture rates on ($^{49}$Ti, $^{51}$Ti, $^{52}$Ti, $^{53}$Ti, and $^{54}$Ti) as function of density for different selected temperatures. Densities are in units of g-cm$^{-3}$. Temperatures are measured in 10$^9$ K and log (E-CAP) represents the log of electron capture rates in units of Sec$^{-1}$.

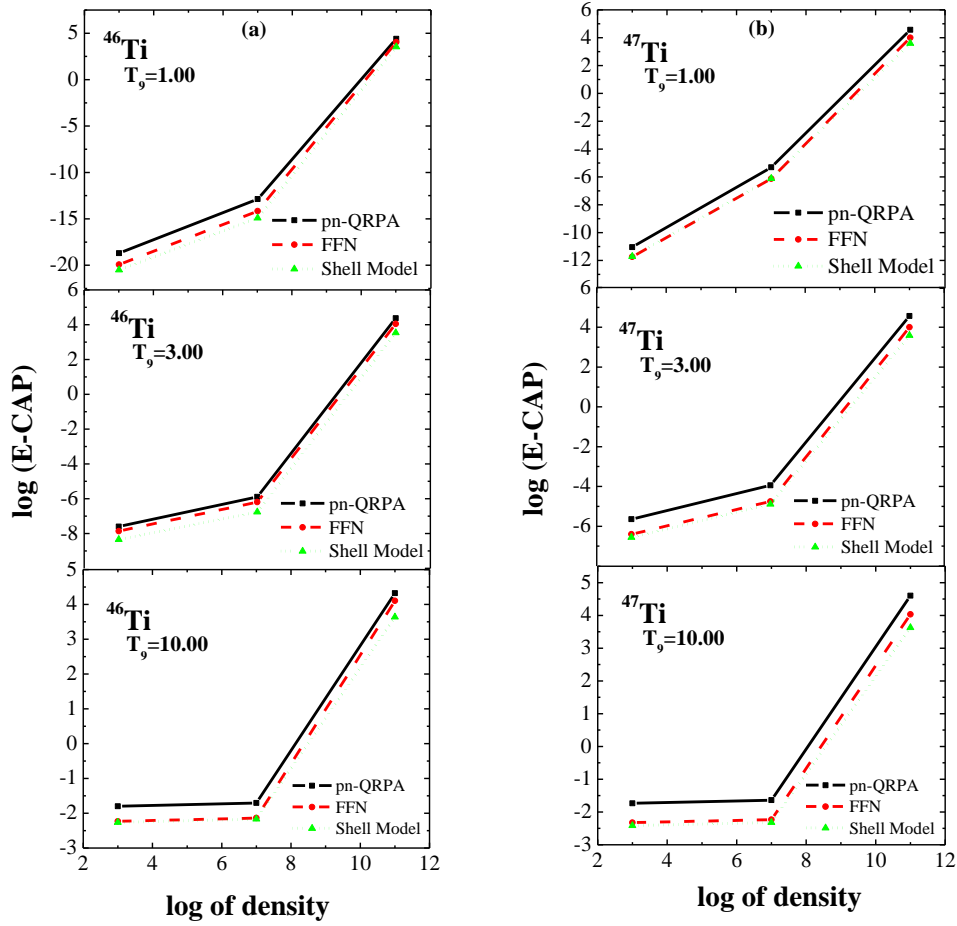

**Fig. 3.** Comparison of electron capture rates (this work) with those of FFN and shell model rates. Solid line represent the electron capture rates of (this work) while dashed line represent the electron capture rates FFN and doted line represent the shell model rates. Log of density has the units of g. cm$^{-3}$ and log (E-Cap) represents the log of electron capture rates in units of Sec$^{-1}$. The temperature $T_9$ measures the temperature in $10^9$K.

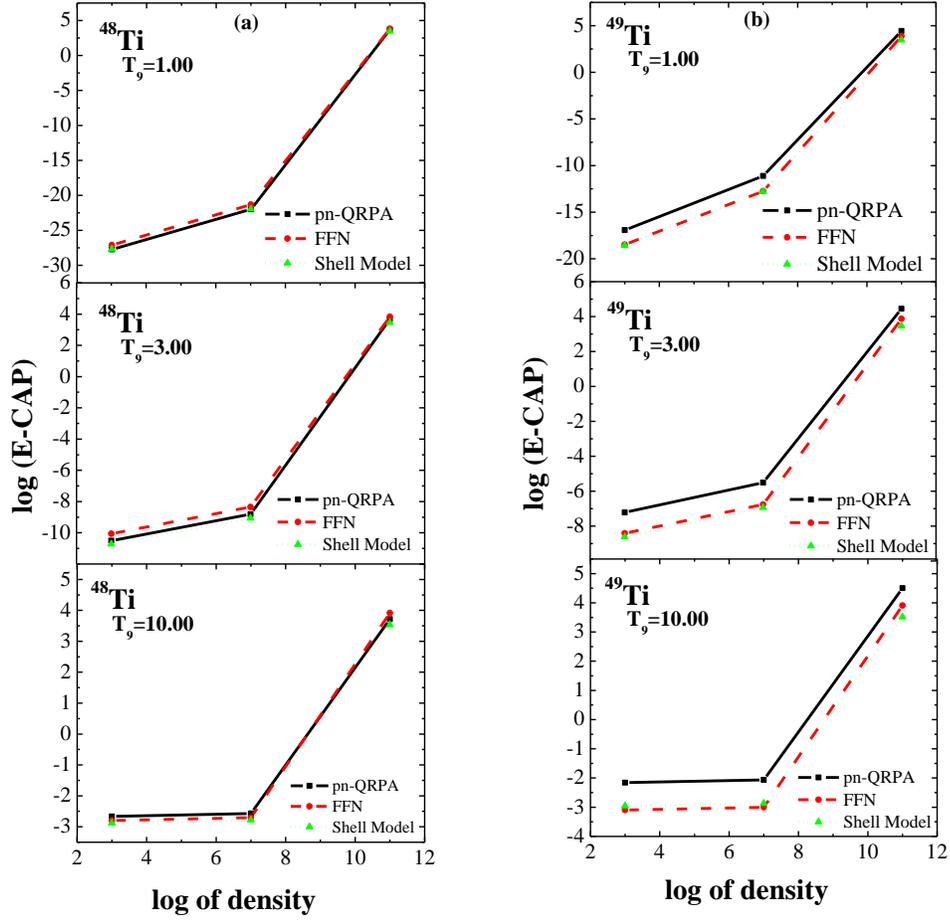

**Fig. 4.** Comparison of electron capture rates (this work) with those of FFN and shell model rates. Solid line represent the electron capture rates of (this work) while dashed line represent the electron capture rates FFN and doted line represent the shell model rates. Log of density has the units of g. cm$^{-3}$ and log (E-Cap) represents the log of electron capture rates in units of Sec$^{-1}$. The temperature $T_9$ measures the temperature in $10^9$K.

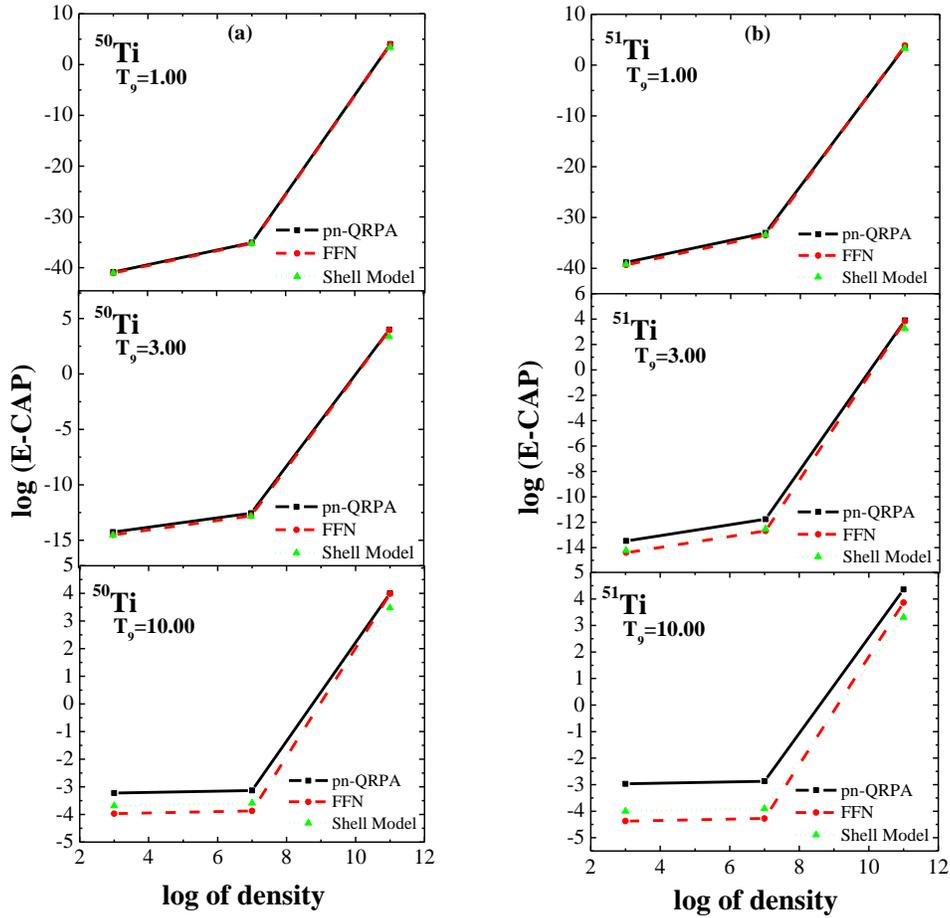

**Fig. 5.** Comparison of electron capture rates (this work) with those of FFN and shell model rates. Solid line represent the electron capture rates of (this work) while dashed line represent the electron capture rates FFN and doted line represent the shell model rates. Log of density has the units of g. cm$^{-3}$ and log (E-Cap) represents the log of electron capture rates in units of Sec$^{-1}$. The temperature $T_9$ measures the temperature in $10^9$K.

This work is partially supported by the ICTP (Italy) through the OEA-project-Prj-16

# Figure captions

**Fig. 1.** Gamow-Teller (GT) strength distributions for $^{48}$Ti. The upper panel shows our results of GT strength for the ground state. The middle and lower panels show the results for the corresponding state measured and calculated by [22] and [23] respectively. $E_j$ represents daughter states.

**Fig. 2** Electron capture rates on ($^{49}$Ti, $^{51}$Ti, $^{52}$Ti, $^{53}$Ti, and $^{54}$Ti) as function of density for different selected temperatures. Densities are in units of g-cm$^{-3}$. Temperatures are measured in $10^9$ K and log (EC) represents the log of electron capture rates in units of Sec$^{-1}$.

**Fig. 3-5** Comparison of electron capture rates (this work) with those of FFN and shell model rates. Solid line represent the electron capture rates of (this work) while dashed line represent the electron capture rates FFN and doted line represent the shell model rates. Log of density has the units of g. cm$^{-3}$ and log (E-Cap) represents the log of electron capture rates in units of Sec$^{-1}$. The temperature $T_9$ measures the temperature in $10^9$K.

# Table captions

**Table I** Calculated electron capture rates on titanium isotopes ($^{40}$Ti -$^{61}$Ti) for different selected densities and temperatures in stellar matter. log $\rho Y_e$ has the units of g-cm$^{-3}$, where $\rho$ is the baryon density and $Y_e$ is the ratio of the electron number to the baryon number. Temperatures ($T_9$) are measured in $10^9$ K. The calculated electron capture rates are tabulated in $\log_{10}\lambda_{ec}$ in units of sec$^{-1}$. In the table, -100 means that the rate (or the probability) is smaller than $10^{-100}$.

# Explanation of Tables and Graphs

**log $\rho Y_e$ (g cm$^{-3}$)**     Where $\rho$ is the baryon density and $Y_e$ is the ratio of the electron number to the baryon number.

**$T_9$**     Temperature in units of $10^9$K.

**E-Cap**     Electron capture rate (s$^{-1}$).